\documentclass{ica}
\usepackage[round,authoryear,comma]{natbib}
\usepackage{mathptmx}
\usepackage[T1]{fontenc}
\usepackage{subfigure}
\usepackage{setspace}
\usepackage{geometry}
\usepackage{epstopdf}
\usepackage{xcolor}
\usepackage{stfloats}
\usepackage{lastpage}
\usepackage[labelsep=period]{caption}
\usepackage{fancyhdr}

\fancypagestyle{first}{
    \fancyhead[L]{\includegraphics[height=1.19cm]{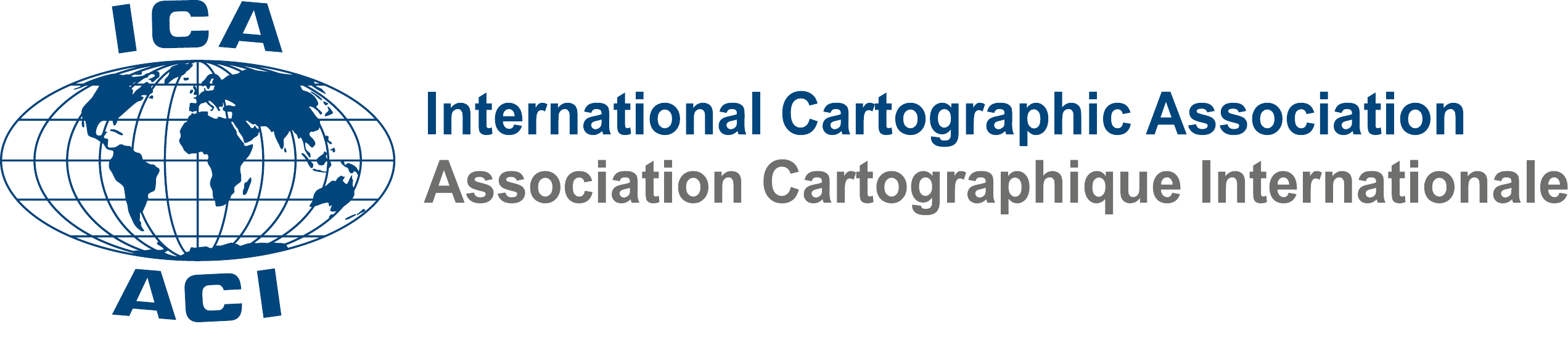}}
    \fancyhead[R]{}
    
}

\begin{document}
\title{Envisioning Generative Artificial Intelligence in Cartography and Mapmaking}

\author{
Yuhao Kang\textsuperscript{a,}\thanks{Corresponding author} , Chenglong Wang\textsuperscript{b}}

\address{
      \textsuperscript{a }GISense Lab, Department of Geography and the Environment, The University of Texas at Austin - yuhao.kang@austin.utexas.edu\\
      \textsuperscript{b }School of Urban Planning and Design, Peking University Shenzhen Graduate School - chenglongw@stu.pku.edu.cn\\
}

\abstract{Generative artificial intelligence (GenAI), including large language models, diffusion-based image generation models, and GenAI agents, has provided new opportunities for advancements in mapping and cartography. Due to their characteristics—world knowledge \& generalizability, artistic style \& creativity, and multimodal integration, we envision that GenAI may benefit a variety of cartographic design decisions, from mapmaking (e.g., conceptualization, data preparation, map design, and map evaluation) to map use (such as map reading, interpretation, and analysis). This paper discusses several important topics regarding why and how GenAI benefits cartography with case studies including symbolization, map evaluation, and map reading. Despite its unprecedented potential, we identify key scenarios where GenAI may not be suitable, such as tasks that require a deep understanding of cartographic knowledge or prioritize precision and reliability. We also emphasize the need to consider ethical and social implications—such as concerns related to hallucination, reproducibility, bias, copyright, and explainability. This work lays the foundation for further exploration and provides a roadmap for future research at the intersection of GenAI and cartography.}


\maketitle

\thispagestyle{first}

\section{Introduction}

Maps have long shaped human understanding of space and place, and are recognized as one of the most important geographic ideas that have changed the world \citep{hanson1997ten}. Cartography, the art and science of mapmaking, has a profound history \citep{krygier1995cartography}. From navigation to visually appealing maps and user-interactive maps, its evolution has always been intertwined with societal demands and technical progress \citep{bagrow2017history}. Today, the needs for personalized, efficient, and creative mapmaking are increasing, and the emergence of generative artificial intelligence (GenAI) presents novel opportunities to fulfill these emerging demands \citep{ai2022some,robinson2017geospatial,sieber2006public}.

GenAI, represented by notable models such as ChatGPT\footnote{https://chatgpt.com/}, Gemini\footnote{https://gemini.google.com/}, and LLaMA\footnote{https://www.llama.com/}, has attracted significant public attention in the recent years. GenAI represents an emerging class of advanced artificial intelligence (AI) methodologies designed to generate new content, such as text, images, or audio, by leveraging patterns and representations learned from existing data \citep{sengar2024generative}. As the field evolves rapidly, there are primary three types of GenAI methods including large language models (LLMs) \citep{zhao2023survey}, diffusion-based image generation models \citep{croitoru2023diffusion}, and GenAI agents \citep{durante2024agent}. They have demonstrated superb capabilities including a comprehensive understanding of world knowledge, high generalizability across a variety of tasks, the ability to handle aesthetics and foster creativity, and facilitating multimodal integration. Collectively, these characteristics provide a wide range of opportunities across diverse domains, including cartography.

\begin{figure*}[t]
\centering
\includegraphics[width=0.92\textwidth]{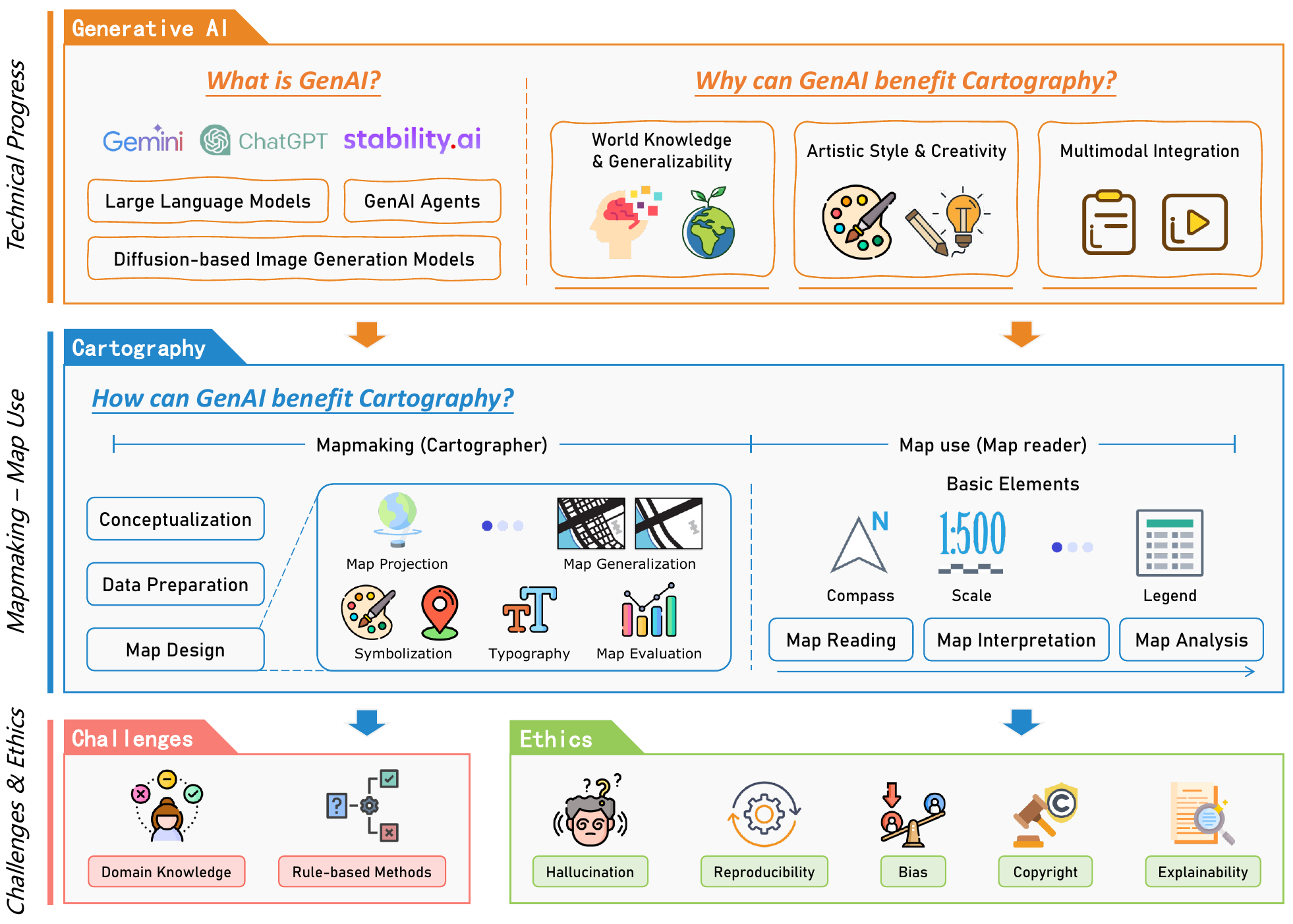}
\caption{A conceptual framework of GenAI for cartography.}
\label{fig:framework}
\end{figure*}

As illustrated in Figure \ref{fig:framework}, building on these technological breakthroughs and increased societal demands, this paper explores several promising directions of GenAI in cartography. Focusing on both mapmaking and map use, we examine the feasibility of GenAI through several case studies: style sheet design, map icon design, map evaluation, and map reading. These examples showcase the unprecedented potential of GenAI to enhance cartographic practices and expand the boundaries of traditional workflows. Moreover, as cartographers, we further delve into key ethical considerations and social implications toward ethical mapmaking.

\section{A historical view of Cartography and Technology}\label{sec:s2}

Cartography, the art and science of making maps, has a history spanning over a thousand years \citep{krygier1995cartography}. From prehistoric cave drawings, such as those found in the Abauntz Cave, Spain \citep{utrilla2009palaeolithic}, to modern printed and digital maps, people have relied on visual representations to convey complex geographic information and facilitate spatial thinking \citep{maceachren1995maps}. The evolution of cartography as a discipline has always been intertwined with societal demands and technological advancements \citep{bagrow2017history}.

During the Age of Discovery, societal demands for exploration and trade led to a focus on creating accurate paper maps using techniques from surveying and geodesy. These maps were crucial for navigation, enabling adventurers and merchants to explore unfamiliar territories. Technological progress in surveying instruments and geodesic methods greatly enhanced the accuracy of these early maps.

With the advent of the printing press, societal demands shifted towards a wider dissemination of geographic knowledge. Cartographers emphasized map design, focusing on both aesthetics and usability in printed maps to reach a broader audience. Technological advancements in printing enabled mass map products and widespread of geographic information across social classes.

In the digital era, societal demands for efficient mapmaking and spatial analysis such as land use mapping drove significant advancements in cartography with the emergence of technologies like Geographic Information System (GIS). GIS transformed the mapmaking process by introducing automation through computer-aided cartography \citep{jones2014geographical}. This significantly reduced the time and effort required by cartographers compared to traditional manual methods. Beyond automation, GIS also enabled users to analyze and visualize geographic data with greater efficiency and precision, facilitating more informed decision-making in a wide range of applications, such as flood mapping \citep{brivio2002integration}, crime mapping \citep{chainey2013gis}, and environmental management \citep{goodchild2003geographic}, etc.

The development of interactive maps later emerged to respond to the societal need for real-time, user-centered exploration of geographic data \citep{andrienko1999interactive,roth2013interactive}. Interactive maps and story maps allowed users to explore geovisualizations in real time, improving engagement and understanding of geographic information. Advances in internet technologies, web development, and user interface and experience design have facilitated these interactive experiences, providing a new level of accessibility and interactivity for end-users.

Today, the societal demand for GenAI in cartography originates from the growing need for more personalized, efficient, and creative mapmaking processes \citep{ai2022some,robinson2017geospatial,sieber2006public}. Users increasingly expect maps that can adapt to their specific needs, whether for planning, education, or storytelling. GenAI may help automate and even personalize map creation, integrate diverse data sources, and produce unique visual styles to meet these demands. Furthermore, there is an increased demand for greater interactivity and user-centered design in cartography, where users can generate their own maps or modify existing ones to deliver their own ideas. GenAI may offer flexibility, efficiency, and creativity in the cartographic process. Given these new opportunities, the following sections will discuss several potential directions of cartography in the era of GenAI.

\section{What is GenAI?}\label{sec:s3}

GenAI refers to an advanced set of AI models and techniques designed to model the underlying patterns and structures of data to generate new content, such as text, images, audio, video, or other modalities \citep{sengar2024generative}. Unlike traditional discriminative models, which primarily focus on categorizing or depicting data, GenAI emphasizes the creation of (entirely) new outputs based on learned representations of data.

The field of GenAI has seen significant breakthroughs over the years. In 2014, innovations such as variational autoencoders and generative adversarial networks (GANs) enabled the generation of images \citep{kingma2013auto,goodfellow2014generative}. Cartographers have also explored the potential of using GANs for map style transfer \citep{kang2019transferring,christophe2022neural}. By the 2020s, further advancements, including diffusion models and transformer-based deep neural networks, made it possible to create high-quality paintings and textual content \citep{vaswani2017attention}. Cartographers have also started leveraging these approaches for map generalization \citep{feng2023polygontranslator}. Today, GenAI is experiencing a significant increase in attention and development. Here, we identify and categorize three types of GenAI models that can play important roles in cartography.

\textbf{LLMs and their variants:} This type of GenAI is designed to capture comlex language patterns from diverse corpora. LLMs have performed exceptionally in tasks such as translation, summarization, question answering, and code completion \citep{zhao2023survey}. Notable models include OpenAI’s GPT series, Google’s Gemini, and Meta’s LLaMA. It is important to note that LLMs are not limited to handling a single modality (i.g., texts). Variants like multimodal LLMs expand these capabilities to convert multiple modalities. For example, OpenAI’s GPT-4 Vision can process both textual and visual inputs, enabling applications such as image captioning and visual question answering, thereby broadening their practical applications. Cartographers have explored the possibilities to leverage LLMs to generate maps \citep{tao2023mapping} or read maps to convert image-based maps to texts \citep{xu2024map}.

\textbf{Diffusion-based image generation models:} These GenAI models could generate high-quality, diverse images from various inputs, such as text descriptions and sketches. Inspired by the concept of diffusion in physics, these diffusion-based models progressively add noise to an image over time and then learn to reverse the noising process to generate realistic images \citep{croitoru2023diffusion}. Several notable models include DALL-E\footnote{https://openai.com/index/dall-e-3/}, Stable Diffusion\footnote{https://stability.ai/}, and Imagen\footnote{https://imagen.research.google/}, which have demonstrated state-of-the-art performance in text-to-image generation. These models open up new possibilities in creative design, visualization, and content creation, bringing our vivid imagination to reality. Cartographers have tried to use diffusion-based GenAI methods to create a variety style of maps \citep{kang2023ethics}.

\textbf{GenAI agents:} GenAI agents, powered by LLMs and their variants, are autonomous AI systems capable of perceiving their environment, reasoning based on the gathered information, and taking actions to achieve specific goals by models themselves \citep{durante2024agent}. By enabling multiple GenAI agents to interact, we may simulate various collaborative workflows and address complex tasks. Cartographers have developed several agents to aid in mapmaking process and cartographic artistic design \citep{zhang2024mapgpt,li2023autonomous,wang2025cartoagent}. These agents communicate through unified, language-based exchanges, with solutions emerging from their multi-turn dialogues.

It should be noted that GenAI is still a rapidly growing field, and there might be more subtypes of GenAI beyond these three primary categories as time evolves. Each category of GenAI methods has its own characteristics and may be suitable for different cartographic tasks.

\section{Why can GenAI benefit Cartography?}\label{sec:s4}

There is no doubt that GenAI provides new possibilities for cartography. Before we jump into specific cases, this section discusses several characteristics of GenAI to understand what are its advantages and how can it suit for different cartographic subtasks. Here are some unique characteristics that GenAI brings, which may not be accomplished with prior statistical and traditional AI methods.

\textbf{World knowledge \& generalizability:} GenAI has the ability to understand huge volumes of facts, concepts, events, and the relationships among entities in the real world altogether. For example, GenAI may associate map elements with their real-world semantic meanings, enabling cartographic design decisions that are contextually appropriate and logically coherent. This ability may facilitate the alignment of the resulting designs with established cartographic principles and common-sense understanding, improving the accuracy and usability of the maps. This is due to its world knowledge capability that allows for associating different aspects and variables together. Furthermore, unlike traditional methods, which are often task-specific \citep{feng2019learning,wang2024transmi}, GenAI is good at handling multiple tasks and may be applied to various cartographic design decisions simultaneously. For instance, a single GenAI model may perform tasks such as reading geographic data, designing map styles, and critiquing map designs, all within a unified framework \citep{wang2025cartoagent}. Traditionally, these steps are separated steps, each requiring specific tools and workflows. GenAI, however, has the potential to handle all of these tasks together, thus enhancing efficiency and demonstrating its high generalizability.

\textbf{Artistic style \& creativity:} One significant advantage of GenAI refers to its ability to generate diverse artistic styles, unlocking human subjectivity and creativity. Traditional cartographic methods often adhere to pre-defined styles and established conventions, which limit their ability to incorporate subjective and artistic perspectives and meet diverse user needs. In contrast, GenAI provides the flexibility to customize map styles to suit specific user preferences or purposes—whether for scientific visualization, public engagement, or storytelling. For instance, a GenAI-generated map could be designed to evoke the aesthetic of a specific historical period, reflecting the styles of that era. Additionally, the use of GenAI may empower the general public, even those with limited cartographic design expertise, to create visually appealing maps, democratizing access to high-quality mapmaking tools. This focus on creativity enhances interactivity, fosters user-centered design, and promotes greater user engagement.

\textbf{Multimodal integration:} Another major advantage of GenAI refers to its ability to process multiple data types and produce a variety of outputs. Traditional methods often focusing on a single type of data input or output. GenAI, however, may smoothly integrate diverse forms of data—such as satellite imagery, geographic data, textual descriptions, and even audio—and produce integrated visualizations. This multimodal capability is particularly valuable in cartography, where representing complex geographic phenomena often requires combining various sources and types of information to create a comprehensive representation. Moveover, this capability enhances the accessibility of cartography. For instance, GenAI may assist blind or visually impaired individuals in accessing and understanding maps by translating visually-represented geographic information into auditory descriptions \citep{robinson2024using}.

\section{How can GenAI benefit Cartography?}\label{sec:s5}

In this section, we explore several potential applications to illustrate how GenAI may benefit cartography. In Section 5.1, we first review several key components of cartography to provide a theoretical framework. In Section 5.2, we introduce potential applications of GenAI in cartography through case studies.

\subsection{Theoretical components of Cartography}

Cartography aims to effectively and accurately convey geographic information between cartographers and map readers through the use of cartographic languages, such as symbols and labels \citep{griffin2021cartography}. Following 
the \textit{Cartography and Visualization} section of the GIS\&T Body of Knowledge \citep{ucgis_cartography_visualization}, this process can be broadly divided into two key components: \textbf{mapmaking} --- the steps carried out by the cartographer, and \textbf{map use} --- centers on how readers interact with maps (Figure \ref{fig:framework}).

\begin{figure}[t]
\includegraphics[width=\columnwidth]{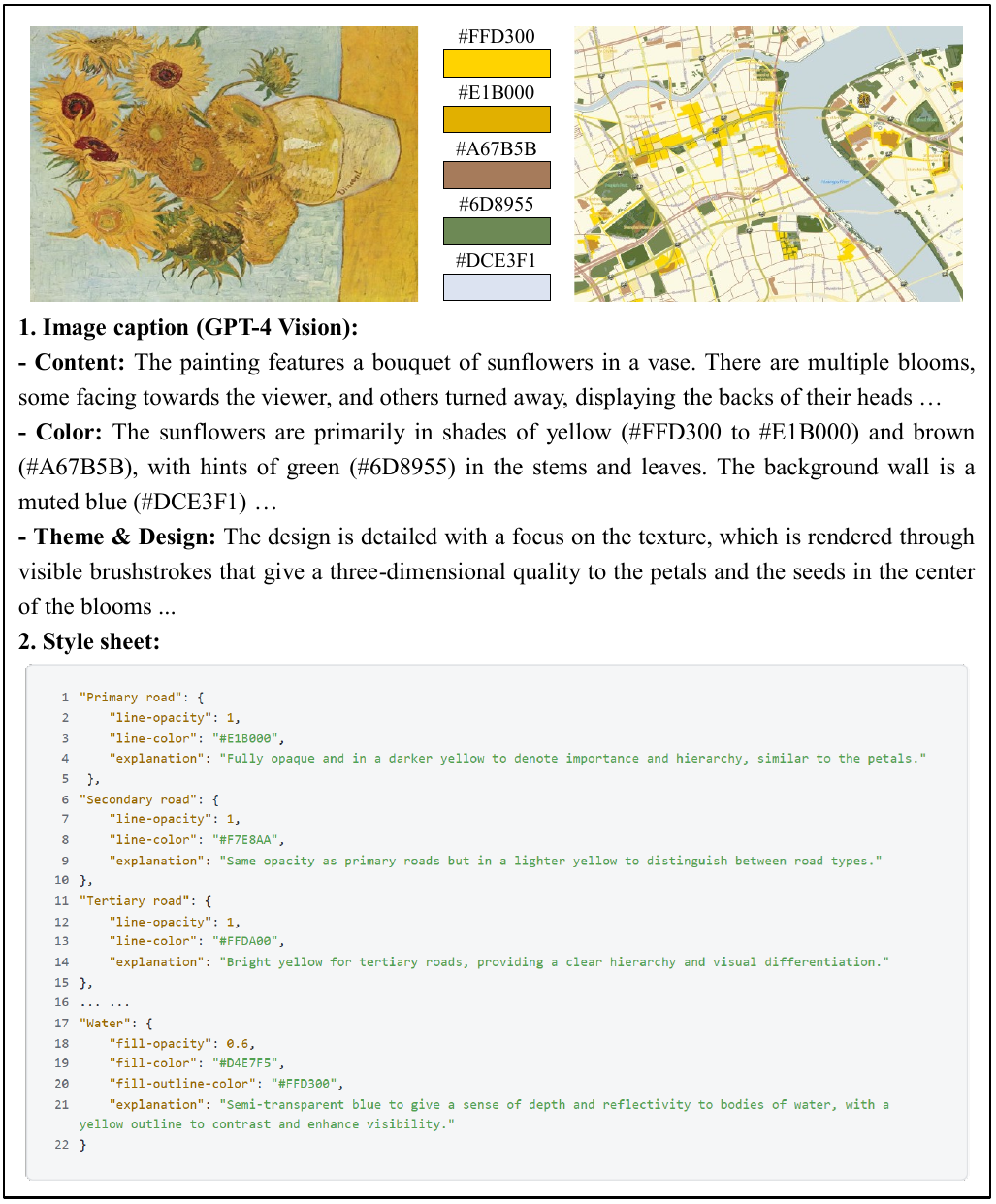}
\caption{Potential applications of GenAI in style sheet design.}
\label{fig:style}
\end{figure}

\textbf{Mapmaking} begins with the \textit{conceptualization} of the map, where its purpose, target audience, and the geographic phenomena to be represented are determined. Once the concept is clear, the process moves to \textit{data preparation}, which includes collecting geographic objects and associated attributes. The next phase refers to \textit{map design}, involving cartographic design decisions made through several interconnected steps. Key aspects include \textit{map projection}, \textit{generalization}, \textit{symbolization}, \textit{typography}, and \textit{evaluation}. \textit{Map projection} involves selecting an appropriate projection to preserve specific properties—such as shape, area, distance, or direction—based on the map’s purpose and audience \citep{battersby2017projection}. \textit{Map generalization} simplifies geographic information to suit the map’s scale, ensuring clarity and legibility while maintaining essential elements \citep{raposo2017generalization}. \textit{Symbolization} translates real-world features into simplified, abstract forms using visual variables like shape, size, color, orientation, and texture \citep{white2017symbolization}. \textit{Typography} focuses on the design, placement, and styling of textual elements on the map \citep{guidero2017typography}. \textit{Map evaluation} is a critical, iterative process that assesses the clarity, accuracy, usability, and aesthetic quality of the map \citep{buttenfield1993representing,kronenfeld2023generalization}. \textbf{Map use}, on the other hand, focuses on the activities of the map reader, including \textit{map reading}, \textit{interpretation}, and \textit{analysis}. \textit{Map reading} is the fundamental step, where users recognize and understand symbols, labels, and features represented on the map \citep{buckley2021reading}. \textit{Map interpretation} goes further by analyzing not only individual map elements but also their relationships \citep{kimerling2016map}. Finally, \textit{map analysis} involves analytical reasoning based on map reading and interpretation. It requires a high level of cognitive ability and prior knowledge to derive insights \citep{kang2024artificial}.
The emergence of GenAI provides new opportunities to benefit the abovementioned two key aspects in cartography.

\subsection{Integrating GenAI in Cartography}

GenAI may support automating mapmaking steps, addressing personalized user needs, and enhancing user engagement. 
In this section, we focus on a few potential subfields that may benefit from the integration of GenAI.

\subsubsection{Symbolization—Style sheet design}

Symbolization involves the use of symbols, colors, patterns, and other graphical elements to represent real-world features and their attributes on a map, as pre-defined in a style sheet. One key advantage of GenAI lies in its artistic style and creativity. This raises the question: Can LLMs design visually appealing style sheets to symbolize map elements that also align with their real-world semantic meanings, based on inspiration sources?

As illustrated in Figure \ref{fig:style}, GPT-4 Vision demonstrates its ability to accurately identify the content, color palette, theme, and design of inspiration sources. Furthermore, it effectively symbolizes map elements, such as roads and water bodies, in a way that reflects their real-world connotations. The resulting styled maps are not only aesthetically pleasing but also effectively communicate geographic information, highlighting the potential of GenAI to perform map style transfer and enhance the artistic aspects of cartography \citep{wang2025cartoagent}.

\subsubsection{Symbolization—Map icon design}

\begin{figure}[t]
\includegraphics[width=\columnwidth]{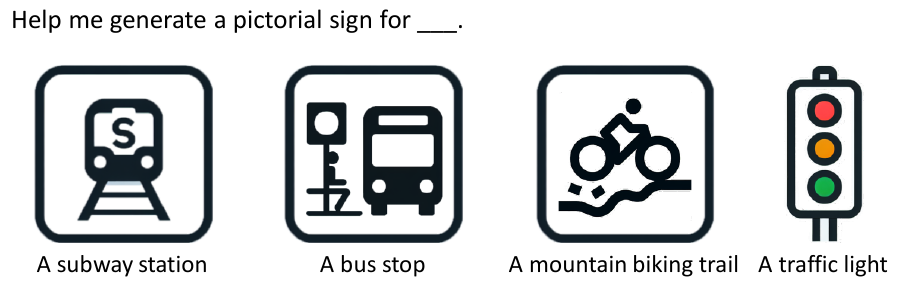}
\caption{Potential applications of GenAI in map icon design.}
\label{fig:icon_design}
\end{figure}

Map icons are clear, easily recognizable markers used to represent map locations. They help users interact with the map intuitively, making it easier to find places. With GenAI, map icons can be custom-designed to meet specific needs and generated efficiently. Traditionally, creating pictorial signs—designs that closely resemble their referents—has been time-consuming. However, as shown in Figure \ref{fig:icon_design}, we can now use tools like DALL-E to quickly generate a series of pictorial signs for various elements such as subway stations, bus stops, mountain biking trails, and traffic lights. These generated icons effectively reflect the semantic characteristics of their referents, making them easy for map readers to understand. In parallel, they follow a minimalism design principle for displaying on maps.

\subsubsection{Map evaluation}

\begin{figure}[t]
\includegraphics[width=\columnwidth]{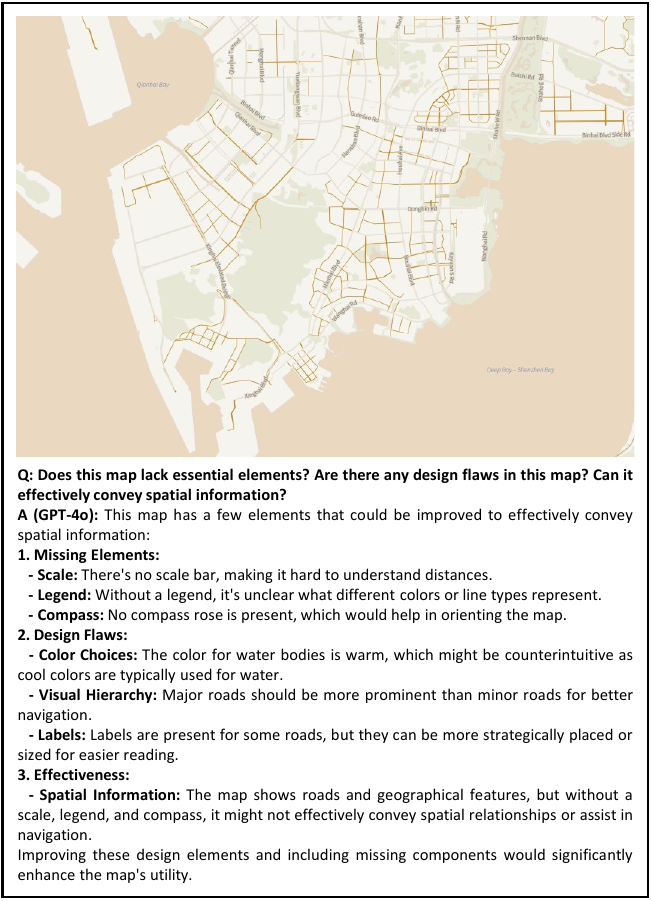}
\caption{Potential applications of GenAI in map evaluation.}
\label{fig:evaluation}
\end{figure}

Map evaluation is an essential component of the mapmaking process, ensuring that maps not only convey geographic information accurately but are also visually appealing. The integration of GenAI into map evaluation like Map Doctor\footnote{https://chatgpt.com/g/g-14WuzVoyJ-map-doctor} has the potential to significantly enhance this process by automating quality assessments, enabling the rapid review of large numbers of maps, and assisting the general public in creating high-quality maps.

As illustrated in Figure \ref{fig:evaluation}, GPT-4o has demonstrated its ability to accurately identify missing map elements (e.g., scale, legend, compass) and detect design flaws. For instance, it identified that the color scheme chosen by the cartographer was suboptimal—warm colors were used for water bodies, which might be counterintuitive since cool colors are typically preferred for representing water. Similarly, it pointed out issues with visual hierarchy, noting that major roads should be more prominent than minor roads to improve map navigability. Additionally, GPT-4o provided suggestions for improving typography, such as strategically placing labels and adjusting their size to enhance readability. These insights are highly valuable, particularly for non-experts, and can greatly enhance both the usability and aesthetic quality of maps.

\subsubsection{Map reading}

\begin{figure}[t]
\includegraphics[width=\columnwidth]{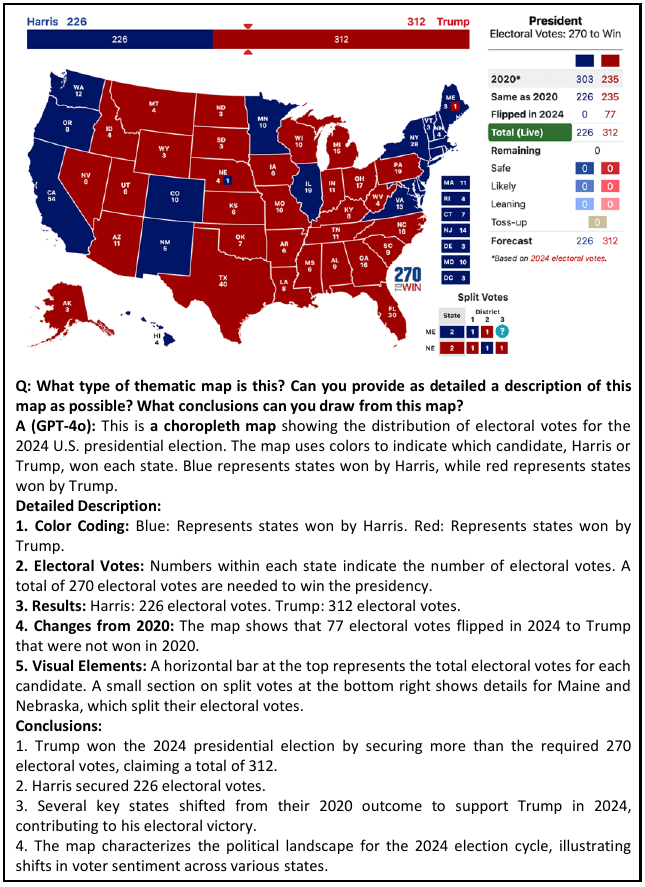}
\caption{Potential applications of GenAI in map reading (choropleth maps).}
\label{fig:reading_choropleth}
\end{figure}

\begin{figure}[t]
\includegraphics[width=\columnwidth]{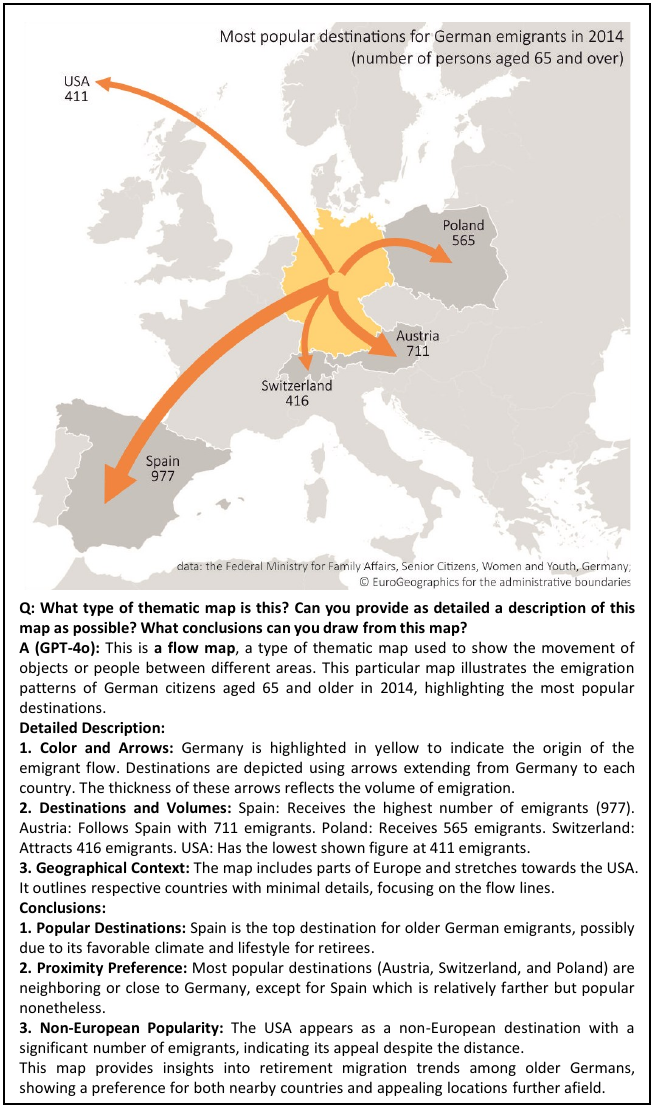}
\caption{Potential applications of GenAI in map reading (flow maps).}
\label{fig:reading_flow}
\end{figure}

The integration of GenAI into map reading offers numerous benefits. First, for the general public, it may enhance user engagement by interactively providing insightful descriptions of maps. Second, in cartographic education, it may simplify complex concepts, making them easier for learners to understand. Third, it may aid blind or visually impaired individuals by translating visually-represented geographic information into alternative modalities, enabling them to access and comprehend maps more effectively \citep{robinson2024using}.

This raises an important question: Can LLMs understand maps? To explore this, we focus on thematic maps, a prevalent type often found online. These maps are commonly used to visualize various natural and social phenomena or events, such as the results of the U.S. presidential election. As illustrated in Figures \ref{fig:reading_choropleth}\footnote{Source: www.270towin.com/2024-election-results-live/president/} and \ref{fig:reading_flow}\footnote{Source: https://gistbok-ltb.ucgis.org/page/24/concept/7223}, GPT-4o demonstrated its ability to accurately recognize and interpret various types of thematic maps, including choropleth maps and flow maps. By leveraging map legends and annotations, GPT-4o provided detailed descriptions and drew valuable conclusions. For instance, in Figure \ref{fig:reading_choropleth}, it correctly identified that Trump won the 2024 presidential election by securing more than the required 270 electoral votes, achieving a total of 312 votes. Moreover, it noted that several key states shifted their support compared to the 2020 election, significantly contributing to Trump’s victory in 2024.

\section{When does GenAI fall short in benefiting Cartography?}\label{sec:s6}

While GenAI offers significant benefits to cartography, it is not without limitations at the current stage. One key limitation of GenAI lies in its lack of domain-specific knowledge. For example, it struggles to fully grasp certain cartographic theories, such as those related to geospatial information visualization like the cartographic cube \citep{maceachren1994visualization} and the information theory of cartography like map communication models \citep{li2020information}. Additionally, mapmaking heavily relies on the rich, intuitive experiences of cartographers, much of which remains difficult to codify or articulate.

Traditional cartographic approaches, such as rule-based methods, still hold significant values and cannot be entirely replaced by GenAI. These methods adhere to pre-defined principles rooted in geographic cartographic theories, ensuring that the generated maps meet strict legal, technical, and aesthetic requirements without unintended variations. Furthermore, they offer transparency and traceability, enabling users and stakeholders to clearly understand the rationale behind each design decision.

As a result, a hybrid approach that combines rule-based methods with GenAI is highly valuable. Rule-based methods provide precision and reliability by strictly adhering to established standards, while GenAI may contribute to creativity, enabling the generation of novel map styles and dynamic responses to user inputs.

\section{Ethics in GenAI for Cartography}\label{sec:s7}
GenAI offers unprecedented potential for cartography. In the meantime, its ethical and social implications should be discussed to ensure the ethical usage in mapmaking and map use. Here, we discuss several key issues including hallucination, reproducibility, bias, copyright, and explainability, each requiring careful consideration in the future GenAI in cartography \citep{kang2023ethics,zhao2021deep,lin2024posthuman}.

\subsection{Inaccuracy due to the hallucination}

GenAI models sometimes generate factually inaccurate, logically inconsistent, or entirely fabricated content—a phenomenon commonly known as \textit{hallucination}. When applied to cartography, hallucination can involve the generation of inaccurate, non-existent, or misleading geographical features on maps. For example, GenAI may create non-existent rivers or mountains, generate fictitious names or labels for locations, or make maps that inaccurately represent scale, distances, or the relative positioning of geographic elements. These factors all pose significant challenges to the scientific accuracy of the maps and can have serious consequences, particularly for users relying on these maps for navigation or decision-making. For example, hallucinated features could lead to planning routes through unsafe or inaccessible areas. In sum, persistent hallucination undoubtedly undermine trust in AI-generated maps and related systems.

To address this issue, researchers may consider the following two potential approaches. On the one hand, improving the model itself is crucial. This can be achieved by using well-curated, up-to-date, and geographically accurate datasets for training, as well as fine-tuning the model with cartographic knowledge. On the other hand, continuous monitoring is also important. Keeping human experts in validation and model development processes may help enhance the accuracy of AI-generated maps.

\subsection{Reproducibility}

Reproducibility, in communities such as GIS and computer science, refers to the ability to achieve consistent results when applying the same method and technique to the same data \citep{wilson2021five}. This raises a question: what does reproducibility mean in the context of cartography?

On the one hand, reproducibility is crucial in the discussion of GenAI for cartography. It allows us to trace model operations, thereby enhancing the explainability of these models. For instance, ensuring the verifiability of outputs is essential for both scientific research and practical applications. On the other hand, placing less emphasis on reproducibility could foster greater creativity and innovation in cartography, allowing practitioners to explore more unconventional designs without being constrained by strict reproducibility standards.

\subsection{Bias}

Bias in GenAI refers to the systematic errors, deviations, or misinformation in a model’s outputs caused by limitations or imbalances in the training data, algorithmic design, or human interpretation. For example, most GenAI models are optimized to maximize the likelihood of reproducing their training data. Consequently, they tend to overrepresent frequent patterns or dominant data distributions while underrepresenting rare or minority patterns, which can have significant negative consequences for certain regions or groups.

For instance, regions with limited geographic data—such as developing countries or sparsely populated areas—may suffer from incomplete or inaccurate maps \citep{lin2024posthuman}. Moreover, maps are not neutral representations of geographic space; rather, they may reflect and reinforce the values, priorities, and intentions of their creators. GenAI systems, which heavily rely on data provided by major technology companies or government entities, often reflect and amplify these dominant perspectives. As a result, their outputs frequently align with certain worldviews and it is crucial to democratize GenAI so that it reflects local and marigized regions’ views. Potential solutions include incorporating data from diverse and underrepresented regions or groups, adopting fairness-aware algorithms that reweight underrepresented data or penalize biased outcomes, and implementing human-in-the-loop AI for cartography. Additionally, fostering community involvement and collaboration with experts in ethics, sociology, and other disciplines is essential to ensure the development of more inclusive and equitable GenAI systems for cartography.

\subsection{Copyright}

Copyright in GenAI is a form of intellectual property that protects content created by GenAI systems, such as the map data utilized for training GenAI systems, GenAI algorithms, and AI-generated map products. The current copyright framework and trademark laws face significant challenges in handling and managing copyright issues. For instance, model developers may unintentionally use map data from official or commercial map providers, potentially exposing themselves to lawsuits for copyright infringement. Another dilemma arises when a map originally designed by a cartographer is input into a GenAI model and subsequently modified based on user-generated prompts. Who, in this case, owns the copyright to the resulting map? Does the copyright belong to the cartographer who created the original design, the user who crafted the prompt, or both?

Addressing these challenges may come from both technical and legal perspectives. Technically, several advanced methods such as machine unlearning could enable GenAI systems to “forget” copyrighted information, thereby possibly mitigating the risk of infringement \citep{nguyen2022survey}. Map service providers and cartographers could also implement technologies like watermarking to protect generated content or develop tools to quantify the contributions of human users in AI-generated works \citep{hosny2024digital}. Legally, further exploration is needed to clarify copyright ownership and establish frameworks that balance the rights and interests of all stakeholders.

\subsection{Explainability}

GenAI models are often referred to as “black boxes” because users may not, or in many cases cannot, fully understand the mechanisms behind the model’s outputs. Explainability, therefore, aims to clarify and interpret the decisions and outputs generated by these models to advance human trust of maps \citep{prestby2023trust}.

The significance of explainability varies depending on the role of the user. For cartographers and GIS professionals, it is essential, as it helps them understand how a GenAI model generates specific maps or arrives at conclusions. GIS professionals rely on the accuracy and validity of the outputs for decision-making in domains like urban planning and resource allocation. More importantly, explainability can provide necessary information when unethical maps are generated. By understanding the logic behind AI-driven processes, cartographers and GIS professionals can trust the results, detect errors or biases, and adjust the models accordingly.

\section{Conclusion}\label{sec:s8}

To conclude, in this paper, we first discuss several unique characteristics that GenAI brings to cartography, including its world knowledge and generalizability, artistic style and creativity enhancement, and multimodal integration. We then explore the opportunities that GenAI offers to benefit a variety of mapmaking and map use tasks. In particular, we provide three example applications including visual variables, map evaluation, and map reading. We also discuss the challenges of integrating GenAI into cartographic practices. Furthermore, we emphasize the importance of ensuring the ethical use of GenAI in mapmaking. Concerns such as hallucination, reproducibility, bias, copyright, and explainability of AI must be acknowledged and addressed.

It is worth noting that our paper may not cover all possible use cases. Also, given the rapid development of GenAI, the examples provided may quickly become outdated with more advanced methods. We believe our study offers valuable contributions for researchers, professionals, policymakers, and the general public to navigate the landscape of potential usage of GenAI in cartography. 
We advocate for more discussions, perspectives, and practices about cartography in the new era of GenAI in the incoming International Cartographic Conference.

\begin{spacing}{1.0}
      \bibliography{ref}
\end{spacing}

\end{document}